\documentclass{bmcart}

\usepackage{amsthm,amsmath}
\usepackage{amssymb}
\RequirePackage[authoryear]{natbib}% uncomment this for author-year bibliography
\RequirePackage{hyperref}
\usepackage{aas_macros_sf} %macros for BibTeX in sans font (used here for literature)
\usepackage[utf8]{inputenc} %unicode support
\usepackage{comment}
\def\includegraphics{}

\startlocaldefs
\endlocaldefs

\begin{document}

%%% Start of article front matter
\begin{frontmatter}

\begin{fmbox}
%\dochead{Research}
\dochead{Meeting report}


\title{A MODEST review}

%%%%%%%%%%%%%%%%%%%%%%%%%%%%%%%%%%%%%%%%%%%%%%
%%                                          %%
%% Enter the authors here                   %%
%%                                          %%
%% Specify information, if available,       %%
%% in the form:                             %%
%%   <key>={<id1>,<id2>}                    %%
%%   <key>=                                 %%
%% Comment or delete the keys which are     %%
%% not used. Repeat \author command as much %%
%% as required.                             %%
%%                                          %%
%%%%%%%%%%%%%%%%%%%%%%%%%%%%%%%%%%%%%%%%%%%%%%

% \author[
%    addressref={aff1}, % id's of addresses, e.g. {aff1,aff2}
%    corref={aff1}, % id of corresponding address, if any
%    noteref={n1}, % id's of article notes, if any
%    email={jane.e.doe@cambridge.co.uk} % email address
% ]{\inits{JE}\fnm{Jane E} \snm{Doe}}
\author[
   addressref={aff_varri},
   email={varri@roe.ac.uk}
]{\inits{A. L.}\fnm{Anna Lisa} \snm{Varri}}
\author[
   addressref={leiden},
   email={cai@strw.leidenuniv.nl}
]{\inits{M. X.}\fnm{Maxwell Xu} \snm{Cai}}
\author[
   addressref={leiden},
   email={fconcha@strw.leidenuniv.nl}
]{\inits{F.}\fnm{Francisca} \snm{Concha-Ramírez}}
\author[
   addressref={aff_dinnbier},
   email={dinnbier@ph1.uni-koeln.de}
]{\inits{F.}\fnm{Franti\v{s}ek} \snm{Dinnbier}}
\author[
   addressref={aff_esa},
   email={nluetzge@cosmos.esa.int}
]{\inits{N.}\fnm{Nora} \snm{L{\"u}tzgendorf}}
\author[
   addressref={aff_pavlik,aff_prgobs},
   email={pavlik@sirrah.troja.mff.cuni.cz}
]{\inits{V.}\fnm{V{\'a}clav} \snm{Pavl{\'i}k}}
\author[
   addressref={aff_sara},
   email={sara.rastello@uniroma1.it}
]{\inits{S.}\fnm{Sara} \snm{Rastello}}
\author[
   addressref={aff_sollima},
   email={antonio.sollima@oabo.inaf.it}
]{\inits{A.}\fnm{Antonio} \snm{Sollima}}
\author[
   addressref={riken},
   email={long.wang@riken.jp}
]{\inits{L.}\fnm{Long} \snm{Wang}}
\author[
   addressref={aff_zocchi1,aff_zocchi2},
   email={alice.zocchi2@unibo.it}
]{\inits{A.}\fnm{Alice} \snm{Zocchi}}

%%%%%%%%%%%%%%%%%%%%%%%%%%%%%%%%%%%%%%%%%%%%%%
%%                                          %%
%% Enter the authors' addresses here        %%
%%                                          %%
%% Repeat \address commands as much as      %%
%% required.                                %%
%%                                          %%
%%%%%%%%%%%%%%%%%%%%%%%%%%%%%%%%%%%%%%%%%%%%%%

% \address[id=aff1]{% unique id
%   \orgname{Department of Zoology, Cambridge}, % university, etc
%   \street{Waterloo Road}, %
%   %\postcode{} % post or zip code
%   \city{London}, % city
%   \cny{UK} % country
% }
\address[id=aff_varri]{%
  \orgname{Institute for Astronomy, University of Edinburgh},
  \street{Royal Observatory, Blackford Hill},
  \postcode{EH9 3HJ}
  \city{Edinburgh},
  \cny{United Kingdom}
}
\address[id=leiden]{%
  \orgname{Leiden Observatory, Leiden University},
  \street{P.O. Box 9513},
  \postcode{2300 RA}
  \city{Leiden},
  \cny{The Netherlands}
}
\address[id=aff_dinnbier]{%
  \orgname{I. Physikalisches Institut, Universit\"{a}t zu K\"{o}ln},
  \street{Z\"{u}lpicher Str. 77},
  \postcode{D-50937}
  \city{K\"{o}ln},
  \cny{Germany}
}
\address[id=aff_esa]{%
  \orgname{European Space Agency, c/o STScI},
  \street{3700 San Martin Drive},
  \postcode{21218}
  \city{Baltimore, MD},
  \cny{United States}
}
\address[id=aff_pavlik]{%
  \orgname{Astronomical Institute of Charles University},
  \street{V~Hole{\v s}ovi{\v c}k{\'a}ch~2},
  \postcode{18000}
  \city{Prague 8},
  \cny{Czech Republic}
}
\address[id=aff_prgobs]{%
  \orgname{Observatory and Planetarium of Prague},
  \street{Kr{\'a}lovsk{\'a} obora~233},
  \postcode{17021}
  \city{Prague 7},
  \cny{Czech Republic}
}
\address[id=aff_sara]{%
  \orgname{Department of Physics, Sapienza-Universit\'a di Roma},
  \street{Piazzale Aldo Moro 5},
  \postcode{00185}
  \city{Roma},
  \cny{Italy}
}
\address[id=aff_sollima]{%
  \orgname{Osservatorio di Astrofisica e Scienza dello Spazio},
  \street{via Gobetti 93/3},
  \postcode{40129},
  \city{Bologna},
  \cny{Italy}
}
\address[id=riken]{%
  \orgname{RIKEN Advanced Institute for Computational Science},
  \street{7-1-26 Minatojima-minami-machi, Chuo-ku}
  \postcode{650-0047}
  \city{Kobe, Hyogo}
  \cny{Japan}
}
\address[id=aff_zocchi1]{%
  \orgname{Dipartimento di Fisica e Astronomia, Universit\`{a} degli Studi di Bologna},
  \street{via Gobetti 93/2},
  \postcode{40127},
  \city{Bologna},
  \cny{Italy}
}
\address[id=aff_zocchi2]{%
  \orgname{Scientific Support Office, Directorate of Science and Robotic Exploration, European Space Research and Technology Centre (ESA/ESTEC)},
  \street{Keplerlaan 1},
  \postcode{2201 AZ},
  \city{Noordwijk},
  \cny{The Netherlands}
}

%%%%%%%%%%%%%%%%%%%%%%%%%%%%%%%%%%%%%%%%%%%%%%
%%                                          %%
%% Enter short notes here                   %%
%%                                          %%
%% Short notes will be after addresses      %%
%% on first page.                           %%
%%                                          %%
%%%%%%%%%%%%%%%%%%%%%%%%%%%%%%%%%%%%%%%%%%%%%%

%\note{All authors are equal contributors.\\
%	(\textbf{Shouldn't we sort the authors by the last names alphabetically?} % Name?
%	\textit{It is already done (C, C-R, D, P, S, V, W, Z), unless there has been a change of alphabet.}) % Vaclav
%    }     % note to the article
%\note[id=n1]{Equal contributor} % note, connected to author

\end{fmbox}% comment this for two column layout

%%%%%%%%%%%%%%%%%%%%%%%%%%%%%%%%%%%%%%%%%%%%%%
%%                                          %%
%% The Abstract begins here                 %%
%%                                          %%
%% Please refer to the Instructions for     %%
%% authors on http://www.biomedcentral.com  %%
%% and include the section headings         %%
%% accordingly for your article type.       %%
%%                                          %%
%%%%%%%%%%%%%%%%%%%%%%%%%%%%%%%%%%%%%%%%%%%%%%

\begin{abstractbox}

\begin{abstract} % abstract
%\parttitle{First part title} %if any
%Text for this section.

%\parttitle{Second part title} %if any
%Text for this section.
We  present an account of the state of the art in the fields explored by the research community invested in ``Modeling and Observing DEnse STellar systems" . For this purpose, we take as a basis { the} activities of the MODEST-17 conference, which was held at Charles University, Prague, in September 2017.
%summarise {\bf the} activities of the MODEST-17 conference (held at Charles University, Prague, in September 2017), which offered an opportunity to the ``Modeling and Observing DEnse STellar system" community to discuss the rich state of the art of the dynamics of collisional stellar systems. 
%We summarise {\bf the} activities of the MODEST-17 conference (held at Charles University, Prague, in September 2017), which offered an opportunity to the ``Modeling and Observing DEnse STellar system" community to discuss the rich state of the art of the dynamics of collisional stellar systems. 
Reviewed topics include recent advances in fundamental stellar dynamics, numerical methods for the solution of the gravitational $N$-body problem, formation and evolution of young and old star clusters and galactic nuclei, their elusive stellar populations, planetary systems, and exotic compact objects, with timely attention to black holes of different classes of mass and their role as sources of gravitational waves.    

Such a breadth of topics reflects the growing role played by collisional stellar dynamics in numerous areas of modern astrophysics. Indeed, in the next decade many revolutionary instruments will enable the derivation of positions and velocities of individual stars in the Milky Way and its satellites, and will detect signals from a range of astrophysical sources in different portions of the electromagnetic and gravitational spectrum, with an unprecedented sensitivity. On the one hand, this wealth of data will allow us to address a number of long-standing open questions in star cluster studies; on the other hand, many unexpected properties of these systems will come to light, stimulating further progress of our understanding of their formation and evolution.

\end{abstract}

%%%%%%%%%%%%%%%%%%%%%%%%%%%%%%%%%%%%%%%%%%%%%%
%%                                          %%
%% The keywords begin here                  %%
%%                                          %%
%% Put each keyword in separate \kwd{}.     %%
%%                                          %%
%%%%%%%%%%%%%%%%%%%%%%%%%%%%%%%%%%%%%%%%%%%%%%


% MSC classifications codes, if any
%\begin{keyword}[class=AMS]
%\kwd[Primary ]{}
%\kwd{}
%\kwd[; secondary ]{}
%\end{keyword}

\end{abstractbox}
%
%\end{fmbox}% uncomment this for two-column layout

\end{frontmatter}

%%%%%%%%%%%%%%%%%%%%%%%%%%%%%%%%%%%%%%%%%%%%%%
%%                                          %%
%% The Main Body begins here                %%
%%                                          %%
%% Please refer to the instructions for     %%
%% authors on:                              %%
%% http://www.biomedcentral.com/info/authors%%
%% and include the section headings         %%
%% accordingly for your article type.       %%
%%                                          %%
%% See the Results and Discussion section   %%
%% for details on how to create sub-sections%%
%%                                          %%
%% use \cite{...} to cite references        %%
%%  \cite{koon} and                         %%
%%  \cite{oreg,khar,zvai,xjon,schn,pond}    %%
%%  \nocite{smith,marg,hunn,advi,koha,mouse}%%
%%                                          %%
%%%%%%%%%%%%%%%%%%%%%%%%%%%%%%%%%%%%%%%%%%%%%%

%%%%%%%%%%%%%%%%%%%%%%%%% start of article main body
% <put your article body there>

% Please note that all names below 
\section*{Main text}
\vspace{0.5cm}
\section{Introduction}
% Anna Lisa Varri
\label{intro}

MODEST, which is an abbreviation for ``Modeling and Observing DEnse STellar systems'', was established in 2002 as a collaboration between groups working throughout the world on the dynamics of dense, collisional stellar systems. Originally, the emphasis was exclusively on theoretical and computational investigations, with the high-level goal of providing ``a comprehensive software framework for large-scale simulations of dense stellar systems, within which existing codes for dynamics, stellar evolution, and hydrodynamics could be easily coupled and compared to reality'' \footnote{The original MODEST mission statement and the description of many initiatives are recorded at  \url{http://www.manybody.org/modest/}}.  Such a vision has quantitatively shaped the activities of the collaboration in the past fifteen years, with several projects pursued by a number of close-knit working groups, during regular small-scale workshops. Many results and tools which have emerged from these initiatives have had a crucial role in defining the current interpretative paradigm for the formation and dynamical evolution of collisional stellar systems.  

The MODEST community has also progressively inspired and enabled the creation of a stimulating arena for a broader discussion on the theoretical understanding and empirical characterisation of several classes of stellar systems, especially globular star clusters and galactic nuclei. Such a broadening of the scope of the collaboration was reflected in a steady growth of the number of participants to its periodic meetings, which have now become large-scale international conferences and have been organized in Europe, North and South America, and Asia.   

The present article follows the trail of the long-standing, yet somehow intermittent, tradition of papers summarising the MODEST activities \citep{Sum_M1,Sum_M2,Sum_M6,Sum_M7}. We, as a group of ten young MODEST members, wish to { present an account of the state of the art in fields explored during} the MODEST-17 conference\footnote{ \url{https://modest17.cuni.cz/}}, which was held at Charles University in Prague, Czech Republic, in September 2017, thanks to the generous hospitality of the Prague Stellar Dynamics Group. The conference represented the primary gathering of the MODEST community in 2017, attended by more than a hundred participants, with a programme structured in nine sessions, spanning from numerical methods for the solution of the gravitational $N$-body problem to the next observational frontiers. Such a wide range of topics demonstrates the growing role of collisional stellar dynamics in a number of astrophysical areas.

The next decade will witness several new instruments which will deliver astrometric, photometric and spectroscopic data of unprecedented sensitivity and accuracy. The recent detection of gravitational waves opens a unique window on the realm of compact objects. Numerical simulations are expected to tackle progressively larger $N$-body systems and to include more realistic approaches to model a variety of multi-scale, multi-physics problems, benefiting  from both software and hardware advances. This wealth of data and tools will allow us to address a number of long-standing open questions in star cluster studies, but also to formulate new ones. Many unexpected properties of these systems are already coming to light, challenging our understanding of their formation and evolution processes.

Despite the richness of the scientific programme\footnote{The electronic version of the book of abstracts, which includes also almost all posters, is available at: \url{https://modest17.cuni.cz/doc/modest17-boa.pdf}} we are reporting on, this summary should not be considered as a complete assessment of the state of the art in collisional stellar dynamics. For a comprehensive view on theoretical and computational aspects we refer the reader to the books by \citet{HeggieHut2003}, \citet{Aarseth2003}, \citet{Merritt2013}, and \cite{AMUSE}. Several review articles are also available on: the internal dynamics of star clusters \citep{Vesperini2010, Heggie2011, McMillan2015}, young massive star clusters \citep{PortegiesZwart2010,Longmore2014}, multiple stellar populations in globular clusters \citep{BastianLardo2018, Gratton2012}, star clusters in merging galaxies \citep{Renaud2018}, as well as the summary of a recent conference on the formation of globular clusters in a cosmological context \citep{Forbes2018}.

\section{Stellar dynamics}
% Francisca Concha-Ramírez
% Anna Lisa Varri
% Frantisek Dinnbier
%edited by ALV
\label{dynamics} 
\subsection{The new phase space complexity of old globular clusters}
\label{dynamics_1}

The study of the internal dynamics of globular clusters (GCs) is traditionally pursued under a relatively stringent set of simplifying assumptions, chiefly isotropy in the velocity space, absence of internal ordered motions, and spherical symmetry. Thanks to new astrometric measurements provided by Gaia (see Section \ref{observations}) and decades-long observational campaigns with the Hubble Space Telescope (HST), supplemented by targeted spectroscopic surveys \citep[e.g., see][]{GES2012, Ferraro2018, Kamann2018}, we are currently undergoing an observational revolution, which is finally starting to reveal to us the phase space properties of many nearby Galactic globular clusters \citep[e.g., see][]{Watkins2015, Helmi2018}. As a result, a new degree of kinematic richness is emerging in this class of stellar systems \citep[e.g., see][]{Bellini2017,Bellini2018,Libralato2018,Bianchini2018}.  Along with this, the availability  of powerful computational tools (see Section \ref{numerical_methods}) calls for theoretical efforts on some forgotten aspects of collisional stellar dynamics, especially regarding the exploration of the role of angular momentum and anisotropy in the velocity space.

In this perspective, some recently addressed topics concern the implications of non-trivial initial conditions for numerical simulations of star clusters, specifically the evolution of spherical isolated systems with primordial anisotropy \citep{Breen17}, and the effects of relaxing the assumptions of synchronization and coplanarity between internal and orbital angular velocity vectors \citep{Tiongco18}. The interplay between the internal evolution of globular clusters and the interaction with the tidal field of their host galaxy also affects the evolution of anisotropy and rotation, generating interesting complexity in the kinematic properties of the clusters \citep{Tiongco17}.

The kinematic properties of the outskirts of globular clusters have also been gaining growing attention. Recent observations show the existence of diffuse, spherical stellar ``envelopes'' surrounding a number of Galactic globular clusters \citep[e.g., see][]{Olszewski2009, Kuzma2018} and extending several times the nominal truncation radius (as estimated on the basis of simple, lowered isothermal models). The assessment of reliable star cluster members and the measurement of their kinematics, both in the plane of the sky and along the line of sight, poses great challenges due to the background confusion limit, but, in the few cases in which it has been accomplished at sufficiently large distances from the cluster centre, it has revealed an apparent lack of a truncation \citep[e.g., see][]{Marino2014}. The nature of such stellar structures is therefore arousing great interest, and several possible formation (and disruption) scenarios have been proposed, including the presence of a population of ``potential escapers'' \citep[see][]{Heggie2001, Kuepper2010, Daniel2017, Claydon2017}, of tidal debris of accreted dwarf galaxies \citep[e.g.][]{Sollima2018}, or stellar structures associated with a possible small dark matter halo \citep[e.g.][Breen et al. in prep.]{Ibata2013, Lee2017,Penarrubia2017}. Such kinematic measurements are made particularly difficult also by the presence of binary stars. Unresolved binaries may determine an inflation of the velocity dispersion profile \citep[e.g.][]{Bianchini2016}, which, in turn, can cause an overestimation of the dynamical mass of the system \citep[e.g.][Moyano Loyola, in prep.]{McConnachie2010}.    

\subsection{Recent progress on some fundamental aspects of collisional dynamics}

In addition to the role played by the emerging ``kinematic richness'', our understanding of several fundamental aspects of collisional stellar dynamics is also far from complete.  
Fresh attention has been recently devoted to the study of the moment of core collapse in idealised $N$-body models. Approaches for a new definition of core collapse based on self-similar solutions have been developed \citep{pavl_subr}. Single component models of core collapse result in radial density profiles similar to those of previous simulations and in agreement with theoretical predictions. Multi-component models, on the other hand, result in time scales for core collapse that correlate with the times of formation of the first hard binaries \citep[see also][]{Fujii2014}. { In this context, we note that  
\citet{Tanikawa2012} demonstrated by means of direct N-body simulations that dynamically formed hard binaries  
usually originate from a strongly interacting group of four or more  
stars. This is in contrast with earlier claims that hard binaries  
originate from triple encounters.}

New characterisations of the processes of relaxation and mixing have also been proposed \citep{Meiron2018}. While ``relaxation time'', as associated with the energy diffusion process, is a useful definition when considering the cluster as a whole, the concept of ``mixedness'' may give a better picture of motion diffusion when particles of particular orbital families are considered.

\subsection{Towards the large \texorpdfstring{$N$}{N} regime}
Recent hardware and software advances \citep{Nitadori2012} have finally allowed us to ``solve'' the gravitational million-body problem for selected globular clusters \citep[see especially][]{Wang2016,Heggie2014}, but several challenges are none the less still present in realistic GC modelling, such as the number of particles moving towards $N=10^7$ and primordial binaries contributing to the extreme time scale differences.
Performing state-of-the-art numerical simulations such as those presented by \citet{Wang2016} still require years of computing time on GPU (Graphics Processing Unit) supercomputers. The development of new algorithms for an efficient treatment of systems at the interface between collisional and collisionless dynamics is therefore urgently needed. Symplectic Particle tree and Algorithmic Regularization Code for Star Clusters (SPARC-SC) is a new particle-particle particle tree code designed for realistic massive star cluster simulations (Wang et al., in prep.). More details about the numerical developments can be found in Section~\ref{numerical_methods}.

Alternatively, ``frozen'' $N$-body models, which follow the dynamics of a tracer particle in the potential generated by $N$ fixed particles, may serve as a tool to study the validity of the continuum collisionless limit ($N \rightarrow \infty$). In this context, some recent results (Di Cintio et al., in prep) show that, consistently with previous work, the orbits evolved in frozen $N$-body potentials qualitatively resemble orbits integrated in the parent, smooth potentials. However, the dependence on $N$ is non-trivial, and, concerning $N$-body chaos, the continuum limit may be questioned. Regular and chaotic orbits for large $N$ models cannot always be safely extrapolated from systems with a lower number of degrees of freedom.

In the purely collisionless regime, a classical topic of current relevance is represented by the characterisation of a gravitational field by means of an expansion over basis functions. A new biorthogonal family of density-potential pairs has been recently proposed by \citet{Lilley2018}, with application to an efficient spectral decomposition of non-spherical potentials. Finally, two recent studies of barred galactic potentials have been conducted, with the identification of a class of three-dimensional non-periodic orbits, which have boxy projections in both their face-on and side-on views \citep{ChavesVelasquez2017}, and the numerical exploration of the threshold for global disc instabilities \citep{ValenciaEnriquez2017}.
 
\subsection{Globular cluster formation in a cosmological context}

The current understanding of GC formation is still only partially satisfactory, both because of several outstanding issues arising from the modelling of the early phases of star formation in a clustered environment and because of the severe computational difficulties in connecting the cluster scale to a cosmological one. 

Given the multi-physics and multi-scale characteristics of this problem, concerted computational efforts are needed to bridge the gap between stellar dynamical and hydrodynamical codes. The software environment AMUSE \citep{Pelupessy2013,PortegiesZwart2013, AMUSE} was indeed developed to fulfil such a need (see Section \ref{numerical_methods}). A novel coupling between AMUSE and FLASH\footnote{\url{http://flash.uchicago.edu/}} \citep{Fryxell2000,2017AAS...22915302W} provides new aspects to the traditional approach of star formation in GCs \citep{McMillanWall2015}. FLASH provides modules for magnetohydrodynamics, bringing feedback from radiation, supernovae, and stellar winds together with the dynamics and stellar evolution modules provided by AMUSE. 

Several key questions that arise in theoretical studies of GC formation and evolution regard their link to the history of our own Galaxy, through the physical mechanisms that shape the population of clusters we see at $z=0$. Could the products of regular cluster formation at high redshift have survived until the present day, and are these relics consistent with the properties of local globular cluster populations? In other words, are GCs old young massive clusters that have survived? To address these open issues, several groups are targeting the challenging problem of the formation of GCs in a cosmological context, with a variety of computational approaches { (e.g., see the recent results by \citealt{Ricotti2016,Renaud2017,Li2017,Li2018,Creasey2018,Carlberg2018}), or within broader galaxy formation projects (e.g., see E-MOSAIC - MOdelling Star cluster population Assembly In Cosmological Simulations within EAGLE, \citealt{Schaye2015,Crain2015,Kruijssen2011,Kruijssen2012}; FIRE - Feedback In Realistic Environments simulations, \citealt{Grudic2017,Kim2018,Hopkins2014}; CosmoGrid, \citealt{Ishiyama2013,Rieder2013})}

\section{Numerical methods}
% Václav Pavlík
% Long Wang
% Frantisek Dinnbier
%heavily edited and completed by ALV
\label{numerical_methods}
\subsection{NBODYX}

NBODY6 is a state-of-the-art direct $N$-body code specifically designed to study the dynamical evolution of  collisional stellar systems \citep[for details, see][]{Aarseth2003}. The current release of the code has been upgraded to include the SSE (Streaming SIMD (Single Instruction Multiple Data) Extensions) and GPU compatibility \citep{Nitadori2012}. 

The equations of motion of the individual stars are integrated by a fourth order Hermite predictor-corrector scheme \citep{Makino1991} with individual time-steps coupled to a Ahmad-Cohen \citep{Ahmad1973,Makino1992} neighbour scheme. The major feature of this code is the implementation of regularisation techniques \citep{Kustaanheimo1965,Aarseth1974,Mikkola1990,Mikkola1993}, which are essential for a proper treatment of close encounters. In principle, multiple methods may be combined to achieve an even higher performance, however, it should be noted that decision-making is the bottleneck of each calculation.  
For realistic modelling of stellar clusters, the code evolves single star and binary evolution from synthetic stellar evolution tracks \citep{Hurley2000,Hurley2002}.

The latest release of the code, NBODY7, contains a post-Newtonian treatment of the force calculation, which means that orbit variations and merging of compact object binaries due to general relativity can be simulated. Over the years, members of the MODEST community have also developed separate branches of the code with prescriptions to include a variety of additional physical ingredients, such as the effects of an arbitrary external tidal field (NBODY6tt\footnote{\url{https://github.com/florentrenaud/nbody6tt}}, \citealt{Renaud2011}) and dynamical friction (NBODY6df\footnote{\url{https://github.com/JamesAPetts/NBODY6df}}, \citealt{Petts2015}). 

In order to perform massive star cluster simulations, hybrid parallelization methods (SIMD + OpenMP, GPU) have recently been implemented \citep{Nitadori2012}. 
With a desktop equipped with state-of-the-art NVIDIA GPUs and $4-8$ cores CPU, the simulations of $N\approx 10^5$ systems are now achievable. In addition, NBODY6++GPU\footnote{\url{https://github.com/nbodyx/Nbody6ppGPU}} adds the MPI support, thus million body-scale simulations become possible \citep[see][]{Wang2016} by using GPU computer clusters.

During the conference, one special session (which has been summarised in a dedicated document\footnote{\url{https://modest17.cuni.cz/doc/nbody_discussion_nup.pdf}}) has been devoted to an open floor discussion about maintaining and further developing NBODY6. Here we only mention that two tools have already been established to encourage the community to actively contribute to the development of the code and the support of the userbase: a wiki page\footnote{\url{https://github.com/nbodyx/Nbody6/wiki}}  and a github repository \footnote{\url{https://github.com/nbodyx}}.

\subsection{\texorpdfstring{P$^3$T}{P3T}}
Most of the dynamical simulations of star clusters in previous studies use the direct $N$-body methods with individual time steps (see previous Section).
With a few hundred CPU cores and a few ten GPUs, realistic modelling of million-body GCs is now achievable \citep{Wang2016},
however, the high computational cost (i.e., approximately half a year per half-mass relaxation time) makes this kind of approach still impractical for general studies, such as specific simulations of rich clusters { (especially in the case of highly concentrated systems)} and broad initial parameter sampling. It is therefore crucial to devise new numerical strategies to successfully tackle the ``post-million body problem'', e.g. by initially targeting the efficient investigation of stellar systems with $N>10^7$.   

{ \citet{Oshino2011} invented and \citet{Iwasawa2015} further explored} the Particle-Particle Particle-Tree (P$^3$T) method, which uses particle trees \citep{Barnes1986} for long-range forces and individual time-step Hermite integrator for short-range forces, based on an Hamiltonian splitting.
They found that the P$^3$T method can be much faster than a pure Hermite integrator.
This result suggests a new direction for the development of the next generation of star cluster simulation codes. By adding regularization methods, the P$^3$T code (e.g., SPARC-SC, which is under development by Wang et al.) may become an efficient alternative code for massive star cluster simulations{, making it possible to reach not only $N>10^7$ particles, but also more realistic stellar densities.}

%----------------------------------------------------------------

\subsection{AMUSE}

AMUSE\footnote{\url{https://github.com/amusecode/amuse}} {\citep[Astrophysical Multipurpose Software Environment;][]{SPZ2009, PortegiesZwart2013, Pelupessy2013}} is a software framework for computational astrophysics. Existing codes from different domains, such as stellar dynamics, stellar evolution, hydrodynamics and radiative transfer can be easily coupled using a high-level python script. It is based on a ``kitchen sink'' model, i.e. a realistic multi-physics and multi-scale simulation can be computed by using dedicated solvers for each physical process or component. AMUSE is not a monolithic code, but it interfaces existing mature numerical codes used by the community as modules to perform the actual calculations. The AMUSE interface handles unit conversions, provides consistent object oriented interfaces, manages the state of the underlying simulation codes and provides transparent distributed computing. Different codes in the same domain are incorporated by using the same interface specification, so that users can change a solver/integrator by changing one line of source code in the high-level python script. This makes the architecture highly modular, and allows users to treat codes as building blocks and to choose the optimal ones according to the problem under consideration. A complete, hands-on introduction to this software environment can be found in \cite{AMUSE}.

\subsection{Monte Carlo, Fokker-Planck and other methods}

A number of approximate schemes have offered very valuable alternatives to the computationally demanding direct $N$-body techniques. Two main Monte Carlo approaches, which
use a statistical method of solving the Fokker-Planck equation,
were originally developed by \cite{Spitzer1971} and \cite{Henon1971}. The latter, so-called ``orbit-averaged method'' has been subsequently improved by \cite{Shapiro1985} and \cite{Stodolkiewicz1986}, and, more recently, \cite{Joshi2000} and \cite{Giersz1998}, respectively. Currently, two efficient numerical implementations rest on this distinguished legacy: CMC \citep[Cluster Monte Carlo,][]{Pattabiraman2013,Rodriguez2016} and the MOCCA\footnote{\url{https://moccacode.net/}} code \cite[MOnte Carlo Cluster simulAtor,][]{Hypki2013}. Both implementations have been widely tested against direct $N$-body techniques, and have been  used to explore the dynamical evolution of collisional systems within large parameter spaces of initial conditions (see the first results of the ``MOCCA SURVEY Database'', by \citealt{Belloni2016}), an endeavour which is still unfeasible with direct methods. In recent years, these rapid, approximate schemes have been intensively exploited in particular to assess the astrophysical properties and local merger rate densities for coalescing binary black holes in GCs (see \citealt{Rodriguez2015,Askar2017}, and Sections \ref{comp_objects}, \ref{grav_waves}). 

Finally, another approach based on the direct numerical integration of the orbit-averaged Fokker-Planck equation in energy and angular momentum space has been essential in the early formulation of the current evolutionary paradigm of collisional systems \citep[see especially][]{Cohn1979, Goodman1983}. Along this line, \cite{vasiliev17} has presented a new scheme for simulating the collisional evolution of spherical isotropic stellar systems based on the one-dimensional Fokker-Planck equation, with the publicly available code PhaseFlow \footnote{PhaseFlow is provided within the Agama library by \cite{agama}, available at \url{https: //github.com/GalacticDynamics-Oxford/Agama}.}. This code implements a high-accuracy finite-element method for the Fokker-Planck equation, and can handle multiple-component systems (optionally with a central black hole, by taking into account loss-cone effects).

\section{Dynamics of planetary systems in star clusters}
% Francisca Concha-Ramírez
% Maxwell Cai
% Long Wang
%heavily edited by ALV
\label{exoplanets}

Modelling planetary systems in star clusters poses many different challenges. Indeed, these systems have very different temporal and spatial scales, which leads to a hierarchical architecture. If one attempts to integrate the coupled system (i.e., coevolving planetary systems in star clusters), the integrator will be forced to use very small time steps (comparable to 1/10 of the orbital period of the shortest period orbits), which essentially stalls the simulation. Monte Carlo approaches can indeed effectively circumvent this difficulty, but the results will strongly depend on the quality of initial sampling. Besides, it is usually not possible to take into account various physical processes present in star clusters, such as stellar evolution, mass segregation, and tidal disruption. Also, the fact that the potential in star clusters is typically lumpy and may vary quickly in time makes it unreliable to impose a static tide to the target planetary system. Planet-planet scattering, a process which can greatly affect the dynamical evolution of a planetary system, should also be considered, since multi-planet systems are common. It is highly challenging to apply analytical treatments to multi-planet systems due to their chaotic nature.

Planetary systems can be affected by a dense stellar environment in two major ways: first, during the planet formation process (Stage 1), protoplanetary disks may be subject to truncation { \citep[e.g.,][]{Vincke2015, 2016MNRAS.457..313P}} due to stellar encounters and/or photoevaporation due to the incident FUV photons from nearby O/B stars {\citep[e.g.,][]{Adams2006, 2013ApJ...774....9A}}. Second, as the disk dissipates (Stage 2), planets are no longer protected by the eccentricity damping mechanism of the disk, and their orbital eccentricities and inclinations can be effectively induced by stellar encounters \citep[e.g.,][]{2009ApJ...697..458S,2013MNRAS.433..867H,2015MNRAS.448..344L,2016ApJ...816...59S,2017MNRAS.470.4337C}. 

Despite the intrinsic challenges, much progress has been made in recent years. At Stage 1,  simulations of disk truncations have been greatly simplified by the use of  empirical recipes generated from heuristic fits of hydrodynamical simulations \citep[e.g.,][]{2014A&A...565A.130B}. First-order simulations that ignore the viscous evolution of protoplanetary disks already yield reasonable agreement with the observed disk-size distribution \citep{2016MNRAS.457..313P}, and a few follow-up studies that take into account the viscous evolution and photoevaporation are currently in progress. 

At Stage 2, pioneer work has been carried out by \cite{2009ApJ...697..458S}. They use KS regularization \citep{Kustaanheimo1965} implemented in NBODY6++ \citep{1999JCoAM.109..407S} to model single-planet systems in dense star clusters. Subsequently, \cite{2013MNRAS.433..867H} investigate multi-planet system in open clusters using Monte Carlo scattering experiments. They use the MERCURY package \citep{1999MNRAS.304..793C} to integrate the long-term secular evolution of planetary systems, and the CHAIN package \citep{Mikkola1993} to handle close stellar encounters. \cite{2015MNRAS.448..344L} employ a similar approach to derive the cross-section for planetary systems interacting with passing stars and binaries. 

A census of free-floating planets in star clusters is presented in \cite{2015MNRAS.453.2759Z}. By using NBODY6, \citet{2016ApJ...816...59S} conclude that dynamical encounters excite the orbital eccentricities of planets to a sufficiently high level such that tidal circularization becomes particularly efficient at the perihelion, which in turn leads to the formation of hot Jupiters. 

Finally, a recent study by \citet{2017MNRAS.470.4337C} points out that external perturbations due to stellar encounters and internal planet-planet scattering are two major mechanisms responsible for the instability of planetary systems in star clusters. Furthermore, their results show that planet ejection is a cumulative process: only $3\%$ of encounters are strong enough to induce the orbital eccentricities of the outermost planets by $\Delta e \ge$0.5, and most encounters only result in minute changes in orbital eccentricities. A follow-up study by \citet{2018MNRAS.474.5114C} argue that field planetary systems bear signatures from their parental clusters. For example, planetary systems with more planets tend to have cooler dynamical temperatures (i.e., lower orbital eccentricities and mutual inclinations), because they likely spend most of their time in the low-density outskirts of the parental clusters. As such, the orbital elements of field planetary systems can be used to constrain their birth environments.

\section{Young star clusters and star forming regions}
% Václav Pavlík
% Francisca Concha-Ramírez
% Frantisek Dinnbier
%edited by ALV
\label{young_clusters}

\subsection{Star forming regions}

While young massive stars disperse and heat the surrounding gas, suppressing further star formation locally, their feedback  might enhance or even trigger another star forming event further away.  Expanding bubbles powered by HII regions, stellar winds or supernovae are able to sweep neutral or molecular gas  over length scales of parsecs or tens of parsecs, collecting the gas into clouds, which cools down, gravitationally 
collapses, forms molecules and eventually stars. This mechanism, ``collect and collapse'', was first proposed by \citet{Elmegreen1978}. This process also propagates star formation: under suitable conditions (mainly on the density), one single event of star formation may trigger a sequence of star forming events. \citet{Whitworth1994} and \citet{Elmegreen1994} provide analytical estimates for luminosity of young stars and density of the interstellar medium (ISM) under which star formation can propagate.

To study how widespread triggered star formation is, it is necessary to distinguish it from spontaneous star formation. 
A strong evidence for triggering is considered to be an observation of two or more star forming regions separated by appropriate time intervals, 
and a morphological structure suggesting that the regions are causally connected. 
Another piece of evidence comes from the comparison of the mass of the observed objects with analytical estimates predicted for observed density, luminosity and age. 
Observations (e.g., \citealt{Deharveng2010,Simpson2012}, Seleznev, in prep.) and numerical simulations \citep[e.g.,][]{Dale2015} suggest that unambiguous identification of triggered star formation suffers from several limitations, mainly the uncertainty in dating young stellar objects and gas displacement due to feedback; 
however they indicate that approximately 1/4 to 1/3 of massive star formation is triggered.

Although the presence of filaments in molecular clouds had been known for decades, it was the Herschel Space Observatory which showed their ubiquity \citep{Andre2015}. 
Detailed maps provided by Herschel revealed several puzzling properties of filaments. 
There exists a critical line mass $M_{\rm{lcr}} \simeq 2 c_{\rm{s}}^2/G$ for a filament of sound speed $c_{\rm{s}}$, above which the filament becomes gravitationally unstable.  Non-star forming filaments with line mass $< M_{\rm{lcr}}$ have been shown to be very common in molecular clouds, with some molecular clouds (e.g., Polaris Flare) being entirely composed of these filaments, and not forming stars at all. Another piece of evidence linking star formation to supercritical filaments stems from the observation of prestellar cores. The majority ($\simeq 70$ \%) of prestellar cores are found to lie within supercritical filaments. An unexpected property of filaments is their almost universal width of $\sim 0.1 \, \mathrm{pc}$ (e.g., see \citealt{Arzoumanian2011}, although see \citealt{Panopoulou2016} for a different point of view). The physical processes responsible for the observed properties of filaments are currently a subject of active research, with three suggested solutions: magnetic fields \citep{Nakamura2008,Seifried2015}, large-scale turbulence \citep{MacLow2004} and global self-gravity \citep{Heitsch2008}.

Supercritical filaments have a low Mach ($c_{\rm{s}} \lesssim \sigma_{\rm{NT}} \lesssim 2 c_{\rm{s}}$) non-thermal velocity component \citep{Arzoumanian2013}. 
Understanding of this component might shed light on dynamical state and perhaps also on the origin of filaments. \citet{Toci2015} suggest that the non-thermal velocity component is caused by small amplitude Alfv\'{e}n waves. However, more recently \citet{DiCintio2018} offer a simpler scenario, without invoking magnetic fields. They propose two models: cold collapse of an initially static filament and a perturbed filament. After initial virialisation, the models assume a state with an almost constant virial ratio throughout several hundreds of free-fall time. 
Both the models agree with the observational data.

\subsection{Young stellar objects and the connection to their natal gaseous structures}

Since prestellar cores form in supercritical filaments, the following questions are immediately prompted: for how long do 
prestellar cores trace the underlying gaseous structure? Which physical mechanism unbinds them? 

With a multiwavelength study of the star forming region surrounding NGC~1333, \citet{Hacar2017} recently found a complicated structure of fibers characterized by a non-trivial topology. At some places, the fibers are intertwined to bundles of filaments. As for the more general case of filaments, supercritical fibers are star forming, while subcritical fibers, which are the majority, are non-star forming. 
The age and evolutionary state of a young stellar object can be estimated from the  slope of their spectral energy distribution. There are four classes of young stellar objects, with ``class 0'' being the youngest, and ``class III'' the most evolved (i.e., weak-line T Tauri stars). While class 0 sources closely trace the gaseous structures, the correlation with gas becomes less pronounced for more evolved sources; class III sources are located randomly with the respect to the gas distribution.

Even low-mass, young stellar objects can drive outflows, which may be sufficient to stop the accreting flow from setting the final mass of the young star. 
Alternatively, \citet{Stutz2016} have proposed the ``slingshot mechanism'', which can also explain the termination of the accretion onto a protostar. In this mechanism, protostars are formed within an oscillating filament. 
The youngest protostars are well embedded within the filament because of their large gaseous envelopes. As the protostars evolve and become more compact, they progressively decouple from gas. When the decoupling occurs near the maximum of oscillation, where the acceleration is the highest, the protostar freely moves away from the filament, as confirmed by recent $N$-body simulations \citep{Boekholt2017}. 

\subsection{Young star clusters}

Approximately 80\% of young stellar objects are found in groups and clusters, with the remaining  20\% being distributed in the field. The observed state of young star clusters can help improving our understanding the process of star formation by providing an empirical assessment of the boundary conditions of such process. This is particularly important in the case of massive stars, the formation of which is still poorly understood.

Two main channels for formation of young star clusters have been proposed: by means of the monolithic collapse of a massive, dense, molecular cloud, or by merging of many smaller subclusters (for a recent review, see e.g. \citealt{Longmore2014}). 
The latter possibility implies that the resulting cluster may retain some memory of the original substructures, at least on a time scale which  depends on the mean density of the region encompassing the subclusters themselves. 
In some cases, the boundary between these two scenarios might be blurred as, in sufficiently dense environments, subclusters can already start merging while star formation is still taking place. 

{Evidence of a hierarchical structure in the stellar component of the 30 Doradus region was reported by \cite{Sabbi2012}.} Interesting results on the nearest young clusters containing massive stars were then obtained within MYStIX \citep[Massive Young Star-Forming Complex Study in Infrared and X-ray][]{Feigelson2013} survey. 
The youngest clusters are often elongated in shape, with typical eccentricities in the range $0.3$ to $0.5$, and they often contain several subclusters. 
Tentative evidence \citep{Kuhn2014} suggests that young clusters then progressively evolve towards configurations that are approximately characterized by spherical symmetry and a more homogeneous density. 
Recent studies have also shown that hierarchical substructures, covering up to three orders of magnitude in surface density, are conspicuous in the Orion A and B molecular clouds \citep{Gutermuth2011,Megeath2016}.
The empirical evidence of the hierarchical nature of Orion A and B also suggests that the precursors of massive clusters may likely be highly hierarchical as well.

The merging scenario of cluster formation has been recently investigated in a number of theoretical studies, with different numerical techniques.  
With direct $N$-body models, {\cite{Fujii2012} have explored the hierarchical formation of young massive clusters via mergers of smaller clusters. They subsequently extended this study with an exploration of the products of mergers of turbulent molecular clouds, modelled with SPH simulations \citep{FujiiPZ2015,Fujii2015,Fujii2016}.} By means of collisional $N$-body models, \citet{Banerjee2015} have set limits on the compactness of an initial configuration which can then form, on the time scale of Myr, smooth, spherically symmetric massive clusters similar to NGC~3603.
By means of collisionless simulations, \citet{Grudic2017} found that subcluster mergers tend to produce a cluster of projected density $\mu \sim R^{-2}$, which is in relatively good agreement with observational studies of massive star clusters,  which have profiles of slope $\mu \sim R^{-2.5}$ to $\mu \sim R^{-3}$. 

Star forming regions have higher binary fraction of low-mass stars (typically by factor of 1.5 to 2), if compared to field stars or ``exposed' star clusters. Thus, binary evolution is important even during the relatively short-lived embedded phase. 
This is mainly due to the fact that binaries are more easily destroyed in denser environments and that the stellar density abruptly decreases as the gas is expelled terminating the embedded phase. Using direct $N$-body simulations to model the  destruction rate of binaries of different semi-major axes, \citet{Kroupa1995} 
found that almost all solar-mass stars are formed in binaries. A similar idea was also used to estimate the initial density in star clusters as a function of the cluster mass \citep{Marks2012}. The binary destruction rate in small clusters is faster than in more massive ones. In principle, the observed binary fraction of young star clusters can therefore offer a possible tool to distinguish between the two cluster formation scenarios mentioned above \citep{Dorval2017}.

If massive stars are formed predominantly at the cluster centre, this can leave an imprint in terms of the degree of primordial mass segregation. 
The knowledge of the birthplace of massive stars within a cluster is valuable for testing different theories of massive star formation (i.e., ``competitive accretion', see \citealt{Bonnell1997}), and even the origin of the initial mass function \citep{Bonnell2007}. Some young clusters (e.g. the Orion Nebula Cluster, ONC, see \citealt{Hillenbrand1997}; NGC~3603 see \citealt{Pang2013}; Westerlund~2, see \citealt{Zeidler2017}) indeed show signs of mass segregation.
Despite recent progress on observational side, this is still a matter of continuing debate because mass segregation can also arise dynamically from initially non-segregated clusters (either due to energy equipartition in the case of lower mass ONC, or due to merging of subclusters in the case of more massive star clusters).

The time scale for formation of a star cluster is another important open question. 
The age of individual stars in an embedded cluster can be determined by comparing observed colour-magnitude diagrams with 
pre-main-sequence evolutionary models. 
However, differential extinction, unresolved binaries and activity of pre-main-sequence stars introduce 
systematic errors leading to large uncertainties in the age determination. As a result, the age spread can range from $\simeq 1 \; \mathrm{Myr}$ up to $10 \; \mathrm{Myr}$,  with a possible gradual increase of the star formation rate over time \citep{Palla2000}. Recently, \citet{Beccari2017}, have detected three distinct pre-main-sequences in the ONC, which is still a very young and embedded cluster (see also Section \ref{stellar_pop}). 

Both increasingly more detailed observations and more sophisticated numerical tools, particularly regarding the interplay between stellar and gas dynamics (see Section \ref{numerical_methods}) and the treatment of realistic feedback from massive stars, are needed in order to clarify the issues described above.

\section{New observational frontiers}
% Alice Zocchi
% Anna Lisa Varri
% Antonio Sollima
%edited by ALV
\label{observations}

\subsection{The (more than a) billion star surveyor: Gaia} 
Among the datasets that are going to be delivered in the upcoming years, the most exciting for the star cluster community is arguably the one provided by the European Space Agency (ESA) astrometric mission Gaia \footnote{\url{https://www.cosmos.esa.int/web/gaia}}. The census of sources for which Gaia is expected to provide phase-space coordinates (positions, parallaxes, proper motions) is estimated to include $\sim 2.5\times10^{9}$ objects ($\sim2.5$ times larger than what originally planned). In addition, other information will be provided, including stellar parameters, metallicity, etc., thus increasing the number of dimensions of the phase-space that we will be able to probe. 

The Gaia-ESO survey\footnote{\url{https://www.gaia-eso.eu/}} \citep{GES2012} was designed to complement Gaia, by providing accurate radial velocity measurements obtained with the GIRAFFE and UVES spectrographs at the European Southern Observatory (ESO) for more than $10^5$ stars in both the Galactic field and star clusters. By considering Gaia and Gaia-ESO data simultaneously, it will be possible to improve the membership determination for stars, and thus to clean up the colour-magnitude diagrams of many nearby clusters. This will improve the determination of many global properties of star clusters and will enable a more detailed study of their internal dynamics.

The astrophysical complexity of these systems has become evident with these accurate data. For example, two kinematic components are identified in the $\gamma^2$ Velorum open cluster \citep{2014A&A...563A..94J} with roughly equal numbers of stars having velocity offset of about 2 km$\,$s$^{-1}$: the population having a broader velocity distribution appears to be younger by 1-2 Myr and less concentrated than the other one. Another example is the detection of substructures in the open clusters Chameleon 1 \citep{2017A&A...601A..97S} and NGC 2264 \citep{2017A&A...599A..23V}, which supports a formation scenario where clusters form from the evolution of multiple substructures rather than from a monolithic collapse (see also Section~\ref{young_clusters}).

The first release of Gaia data (DR1-TGAS, Tycho-Gaia Astrometric Solution) provided positions, proper motions and parallaxes for $\sim2\times 10^{6}$ stars originally included in the Tycho mission in the solar neighbourhood. In spite of the relatively large uncertainties, many interesting results have been obtained from this release and its correlation with other surveys \citep[SDSS, LAMOST, RAVE;][]{2017A&A...598A..58H,2017A&A...605A...1R,2017MNRAS.469L..78M}.
For many open clusters, the mean parallaxes and proper motions have been determined using TGAS data \citep{2017A&A...599A..32V}, and they are generally in very good agreement with the earlier determination based on Hipparcos data (with the exception of the mean parallax for the Pleiades cluster).
One of the main issue which emerged from the early analysis of DR1 data concerned the stellar parameters determination: covariances are apparent among the uncertainties of all parameter. Intense efforts have been devoted by several groups to the study of the correlation in the noise. 

The second data release DR2\footnote{\url{https://www.cosmos.esa.int/web/gaia/dr2
}} has just provided data for a larger sample of stars, with an accuracy surpassing the expectations. DR2 contains positions $(\alpha, \delta)$, G and integrated BP and RP photometric fluxes and magnitudes for all sources (with typical uncertainties $\epsilon_{\rm{G}}\sim0.002$~mag), five-parameter astrometric solutions for all sources with acceptable formal standard errors ($30<\epsilon_{\pi}/\mu$as$<700$; for $>10^9$ stars) and radial velocities for sources brighter than 12 mag (with $\epsilon_{\rm{v}}\sim2$~km/s). Moreover, for stars brighter than $G=17$ mag estimates of the effective temperature and, where possible, line-of-sight extinction are provided. 

The community is currently busy mining and interpreting such a transformative dataset and several demonstration and early scientific results have already been published. As an incomplete list of possible applications, here we mention the quantification of the lumpiness of the local halo \citep{Koppelman2018}, the kinematics of the Galactic disk \citep{Katz2018}, the calibration of the RR Lyrae and Cepheids distance scale (\citealt{Clementini2018}, with implications for the measurement of the Hubble constant, see \citealt{Riess2018}), the distance and orbits of GCs and dwarf galaxies (\citealt{Helmi2018}, with implications for the measurement of the Milky Way mass, see \citealt{Watkins2018}), the detection and characterisation of tidal streams \citep{Malhan2018} and hypervelocity stars \citep{Boubert2018, Marchetti2018, Lennon2018, Hawkins2018}. The quality of DR2 data is high enough to determine the fine structure of the Hertzsprung-Russell diagram of stars in the local neighbourhood and beyond \citep{Babusiaux2018}, an endeavour which is revealing many surprises for stellar population studies \citep[e.g., see][]{Kilic2018, Jao2018, El-Badry2018}. As for the study of the internal kinematics of structures in the Local Group, this has been proved feasible even for individual point sources in M33, M31 and the Large Magellanic Cloud \citep{vanderMarel2018,Vasiliev2018}, paving the way for the characterisation of many other objects, including selected young and old star clusters \citep{Bianchini2018,Milone2018,Kuhn2018}. Beside the primary interest on Galactic stellar populations and dynamics, Gaia data enable to measure exoplanet sizes \citep{Fulton2018}, gravitationally lensed systems \citep{Krone-Martins2018}, and even to test general relativity using positional displacement of Sun and Jovian limb. Moreover, it can also be used to develop alternative astrometric search methods for gravitational waves sources \citep{2017arXiv170706239M}. 

\subsection{The electromagnetic spectrum is not enough: gravitational wave detectors} 

The groundbreaking gravitational waves (GW) detection from Advanced LIGO\footnote{\url{https://www.ligo.org/}} \citep{2016PhRvX...6d1015A} provided the first evidence for massive black holes (BHs) mergers, confirming many predictions of general relativity. 

The continued improvement of sensitivity over time will result in more detections at greater distances. The advent of additional detectors around the world (such as KAGRA\footnote{Kamioka Gravitational Wave Detector, \url{http://gwcenter.icrr.u-tokyo.ac.jp/en/}}, scheduled for late 2018) will result in better angular resolution and better chances for the identification of the electromagnetic counterparts of the detected events. Indeed, the joint detection of a GW event by LIGO and Virgo \citep{2017arXiv170909660T} has provided a much better angular resolution than the detections by LIGO alone. Thanks to the joint collaboration between the different GW observatories it will be also possible to use polarization to break the degeneracy between luminosity distance and inclination. 

New insights in this field are also expected from the forthcoming missions involving GW detectors in space. In this regard, it is noteworthy that LISA Pathfinder\footnote{ \url{http://sci.esa.int/lisa-pathfinder/}}, the first high-quality laboratory launched in December 2015 which conducted high-precision laser interferometric tracking of orbiting bodies in space, surpassed its performance requirements and expectations. LISA \footnote{Laser Interferometer Space Antenna, to be launched in 2034, \url{http://sci.esa.int/lisa/}} will provide information about dynamical systems over a wide range of time, length, and mass scales. The advent of such mission will shift the boundary of GW astrophysics to low-amplitude/large distant events like those produced by Galactic white dwarf binaries (Finn et al. 2015), extreme mass ratio and supermassive BH inspirals \citep{2008JPhCS.122a2037G}, extra-Galactic binaries in the field and black hole binaries in GCs up to distances of 30 Mpc \citep{2016PhRvL.116w1102S}. 

\subsection{A new eye on the invisible Universe: Lynx} 
A new era is also beginning for X-ray astronomy, thanks to Lynx,\footnote{\url{https://wwwastro.msfc.nasa.gov/lynx/}} a large area, high angular resolution, X-ray mission concept for the next decade. This mission will combine a large gain in collecting area, an angular resolution of 0.5 arcsec, and very high resolution spectroscopy over a large field of view. With its two orders of magnitude leap in sensitivity over Chandra and ATHENA\footnote{ATHENA, Advanced Telescope for High-ENergy Astrophysics, is an ESA mission to be launched in 2028; \url{http://sci.esa.int/athena/}.}, it will provide insights into many different scientific problems. 

One of the main science drivers of Lynx is the determination of the origin of supermassive BHs. It will be indeed possible to discriminate between the two main hypotheses regarding the seeds of these objects: Population III remnants and intermediate-mass black holes. Lynx will have enough sensitivity to directly detect $10^4$ M$_{\odot}$ objects, pinning down the luminosity function to distinguish between these seeds. In this respect, Lynx would be a useful complement to JWST (James Webb Space Telescope), ATHENA, GMT (Giant Magellan Telescope) and other future missions. 

Lynx is also expected to find and characterize the first generation of groups of Milky Way-sized galaxies around $z \sim 4$. In this regard, it is crucial to investigate the content and ionization fraction of the circumgalactic (CGM) and intergalactic medium (IGM): very different input physics could indeed lead to similar stellar light in galaxies at $z = 0$, but very different CGM and IGM properties. By observing feedback from star formation on scales from individual young star forming regions to the entire galaxy, Lynx will map the IGM, allowing us to uniquely pin down the physics of structure formation. Furthermore, medium-deep Lynx Surveys will expose the emergence of black hole populations in galaxies after $z \sim 6$, in a wide range of galaxy types and density environments.

Moreover, Lynx will provide a detailed view of every aspect of the BH feedback process, by measuring the energy and momentum flux in BH-generated outflows. This will make it possible to study where, how, and how much of the active galactic nuclei (AGN) outburst energy is dissipated in galaxies, groups and clusters.

Finally, with Lynx it will be possible to find and characterize the low $L_{\rm{x}}$ population of GCs. With Chandra, we can probe below few $\times 10^{30}$ erg/s only for a few nearby clusters (M4, 47 Tuc, NGC 6440, NGC 6752, NGC 6397), while with Lynx it will be possible to do this for dozens of Galactic GCs. 

The contribution from the MODEST community to the development of the science cases within the upcoming 2020 Decadal survey has been encouraged. The Lynx observatory will offer a complementary view to dense stellar systems with respect to the facilities illustrated above and will thus be crucial to tackle the many open questions that keep emerging in this era of revolutionary observations.

\section{Stellar populations in star clusters}
% Alice Zocchi
% Anna Lisa Varri
% Antonio Sollima
%edited by ALV
\label{stellar_pop}
\subsection{Initial and present-day mass function}
The original mass distribution of stars, commonly referred to as the initial mass function (IMF), represents one of the central questions in the theory of star formation and has strong relevance for many areas of astrophysics. The universality of the IMF, its shape and the parameters driving its hypothetical variation are still far from being completely understood. 

Some suggestions on possible IMF variations have been recently put forward on the basis of the shape of the present-day MF in unevolved stellar systems ($t_{\rm{age}}<<t_{\rm{rh}}$). While \citet{1998AJ....115..144G} and \citet{2002NewA....7..395W} derived MFs for Draco and Ursa Minor dwarf spheroidals which are consistent with a \citet{1955ApJ...121..161S} IMF, \citet{2013ApJ...771...29G} found evidence of MF variations correlated with the mean metallicity in a sample of ultrafaint dwarf galaxies. This study has been however questioned by \citet{2017MNRAS.468..319E}, who found that evident MF differences cannot be detected with the currently available photometric data. \citet{2013ApJ...762..123W} analysed a large sample of young clusters and associations whose MFs are available in the literature: in spite of the large cluster-to-cluster differences, a careful revision of the associated errors indicates that the hypothesis that they are consistent with a single IMF slope cannot be ruled out. 

Interesting information can be derived also by the study of dynamically evolved stellar systems. The analysis of high-resolution HST photometry for a sample of 35 Galactic GCs revealed tight correlations of the slope of the present-day MF with the half-mass relaxation time and with the fraction of remnants \citep{2017MNRAS.471.3668S}. The observed trends are compatible with the natural concept of dynamical evolution, with highly evolved clusters characterized by flatter MFs and large fractions of remnants, and the small observed spread leaves little room for IMF variation. However, none of the $N$-body simulations run for comparison are able to reproduce the flatness of the measured MFs ($\alpha>-1$ for 27 out 35 analysed GCs) for their relatively low number of elapsed half-mass relaxation times \citep[$3<t_{\rm{age}}/t_{\rm{rh}}<10$;][]{2017MNRAS.472..744B}. Moreover, less evolved GCs ($t_{\rm{age}}/t_{\rm{rh}}<3$) show a depletion of low-mass stars with respect to a single power-law, which should be reminiscent of their IMF shape (as predicted by \citealt{2001MNRAS.322..231K} and \citealt{2003PASP..115..763C}). In a similar comparison between observed MF and $N$-body simulations \citet{2016MNRAS.463.2383W} found that while for the majority of clusters analysed in their sample the present day MF and the degree of mass segregation are consistent with the prediction of simulations starting from a universal IMF, some non-standard initial conditions should be present in NGC 5466 and NGC 6101, which are non-segregated and characterized by a flat MF. The need of IMF variation has been also put forward to interpret the evidence coming from mass-to-light ratios estimated through integrated light analyses of M31 GCs \citep{2017ApJ...839...60H} and the fraction of low-mass X-ray binaries in Virgo GCs and Ultra Compact Dwarf galaxies \citep[UCDs,][]{2012ApJ...747...72D}.

\subsection{Mass-to-light ratios}
An accurate determination of the mass-to-light ratios ($M/L$) of star clusters is important because it provides direct information on their present day stellar population and important clues on their IMF. Understanding the values of $M/L$ from a dynamical point of view is also crucial to determine the evolutionary history of these systems. However, estimates of the $M/L$ from dynamics and from population synthesis do not fully agree, and the effects of several factors need to be considered in order to account for this discrepancy. In particular, the global mass function and the presence of binaries and dark remnants need to be properly assessed, and their determination is particularly tricky because they depend both on the unknown initial conditions and on the dynamical evolution history of the system.

Not only the global $M/L$, but also their radial profiles within the system are crucial to uncover the stellar population content in star clusters. Indeed, mass segregation and the evolution towards a state of partial energy equipartition cause the $M/L$ radial profile not to be constant, and to show a prominent central peak due to the presence of dark remnants \citep{2017MNRAS.469.4359B}. The study of the evolution of numerical simulations show that the exact shape of these profiles depend on the evolutionary stages of clusters. A decrease in $M/L$ is seen for highly evolved systems, and it is primarily driven by the dynamical ejection of dark remnants, rather than by the escape of low-mass star.

A particularly intriguing puzzle is related to the decreasing trend of $M/L_{\rm{K}}$ with metallicity observed in M31 clusters \citep{2011AJ....142....8S}, which cannot be reproduced by means of simple single stellar population models. A proposed explanation for this is related to the role of mass segregation \citep{2015MNRAS.448L..94S}, but this factor alone is not sufficient to completely account for the discrepancy. It has been recently showed that by taking into account dynamical evolution, together with a top heavy IMF, and a 10\% retention fraction for remnants, a better agreement to the data can be obtained \citep{2016ApJ...826...89Z}, and to incorporate the age-metallicity relation further improves the fit \citep{2017ApJ...839...60H}. 

\subsection{Multiple stellar populations}
For decades GCs have been considered as a collection of stars all characterised by the same age and initial chemical composition, and treated as a prototype of a simple stellar population. However, numerous observational studies have put this simple picture to test: spectroscopic observations have unveiled different chemical composition for stars in the same cluster \citep{2004ARA&A..42..385G,2009A&A...505..117C,Gratton2012}, and accurate photometric data obtained with HST have shown the splitting of the evolutionary sequences in the colour-magnitude diagrams (CMDs) of GCs \citep{Gratton2012,2015AJ....149...91P,2015MNRAS.447..927M}. One of the suggested interpretations of these observations is that clusters have been the site of multiple generations of stars with the second-generation (enriched) stars formed from material polluted by the ejecta of some first-generation (pristine) stars.

Several scenarios have been put forward to explain the formation of second-generation stars, each proposing a different source for the polluting material. However, none of these scenarios is able to reproduce the observed detailed chemical properties of clusters, and their origin is thus still unknown \citep{Gratton2012}.
Each one of the formation scenarios proposed so far should have imprinted a typical signature in the spatial distribution and kinematic properties of different populations, even though two-body relaxation determines a progressive mixing \citep{Vesperini2013}. In particular, observations show that enriched stars (i.e., with low O and high Na abundances) are more concentrated \citep{2011A&A...525A.114L} and characterized by a lower fraction of binaries \citep[possibly linked to their high concentration at early epochs;][]{2015A&A...584A..52L} with respect to pristine stars. 

Recently, observational efforts have been dedicated to other systems, such as clusters in nearby galaxies and young stellar clusters (see \citealt{BastianLardo2018}), in order to determine the importance of the environment and of the epoch of formation of these systems in shaping their populations. Theoretical studies have also explored the dynamical properties of such multiple populations, in view of the possibility of using high-quality kinematics to unravel clues about their origin {\citep[e.g.,][]{Vesperini2013,2015MNRAS.450.1164H,Mastrobuono2013,Mastrobuono2016}.}

\citet{2016Natur.529..502L} reported an age difference of a few Myr among stellar populations in three clusters of the Magellanic Clouds (NGC 1783, NGC 1806, and NGC 411). In the case of NGC 411, they have reported a reversed radial behaviour of the two populations with respect to what is observed for the majority of the Galactic GCs, with the enriched population dominating in the outskirts of the cluster. A minor merger (mass ratio less than 0.1) scenario seems to be appealing to reproduce this feature \citep[see also][]{2010ApJ...722L...1C,2013MNRAS.435..809A,2016MNRAS.461.1276G}.
A study of the dynamical properties of the merger remnants \citep{2017MNRAS.472...67H} shows that they are expected to be characterized by some degree of rotation and radial anisotropy; unfortunately, these kinematic properties are still beyond the current observational capabilities. It has been discussed that, by considering a different background subtraction compared to the approach by \citet{2016Natur.529..502L}, \citet{2016MNRAS.459.4218C} found no reversed radial behaviour for the two populations, but such a feature has been detected in other systems \citep[see for example the case of M15,][]{2015ApJ...804...71L}.

Moving on to even younger systems, three discrete sequences have been recently discovered by \citet{Beccari2017} for pre-MS stars in the ONC. Among these populations, the older stars appear to be less concentrated than younger ones, and to rotate more slowly, hinting to a possible slowdown as they evolve. The origin of this discreteness is still uncertain. Several possible explanations have been proposed: the presence of unresolved binaries, rotation, or a genuine age difference. Only in the last case, the three sequences would correspond to different episodes of star formation. 

The empirical characterization of the dynamics of the multiple population phenomenon is arousing progressively greater interest, and, when possible, attention is being devoted to the observational study of the phase space properties of distinct populations \citep[][see also Section~\ref{dynamics_1}]{Bellini2017,Bellini2018,2017MNRAS.465.3515C,Libralato2018, Milone2018}.    
One additional recent example is represented by the case of M54, for which MUSE observations (Alfaro-Cuello et al., in prep.) provided an exceptional sample of kinematic and chemical data, allowing the identification of two distinct stellar populations \citep[see also][]{2008AJ....136.1147B}, with different metallicities, velocity dispersion and possibly rotation profiles. Such an object is indeed particularly complex, not just because it might constitute a ``link'' between classical globulars and nuclear star clusters, but also given its proximity to Sagittarius, which makes the membership assessment particularly challenging (especially regarding possible contamination of the younger population).

\subsection{Stellar population in extra-Galactic systems}

Recently, much efforts have been put also into the analysis of the dynamical and general properties of stellar populations in extra-Galactic environments. Full spectral synthesis analysis of nuclear star clusters (NSCs) in several galaxies reveals generally old ages for their stellar contents ($\sim 10$ Gyr), with some evidence of prolonged star formation until the present day (Kacharov et al., in prep.). However, every NSC appears to be characterised by a complex star formation history (e.g., see the case of NGC 5102 and Section \ref{nuclei}). 

Simple population synthesis has been recently used also to trace back the properties of progenitors of GCs and ultra-compact dwarfs \citep{2017arXiv170807127J}. From these preliminary studies it seems possible that these objects could be as bright as quasars at high redshifts and thus detectable with the upcoming JWST. In this regard, exciting evidence of young ($t_{\rm{age}}<100$ Myr), low-mass ($M<10^{7}$ M$_{\odot}$) and low-metallicity ($Z<0.1$ Z$_{\odot}$) stellar systems compatible with the expected properties of the progenitors of present-day GCs have been recently revealed at redshift $z\sim3.2$ thanks to the high magnification produced by the gravitational lensing of the foreground galaxy cluster MACS J0416 \citep{2017MNRAS.467.4304V}.

Finally, considering GC systems as a whole, differences between the GC content of the Fornax and Virgo galaxy clusters are apparent, with GCs in the Virgo cluster being more concentrated, when observed in $g$-band, than those in Fornax, while the two distributions overlap when near-infrared luminosities are considered (Dabringhausen et al., in prep.).

\section{Galactic centre / nuclear star clusters}
% Sara Rastello
% Long Wang
% Nora Luetzgendorf
%heavily edited by ALV
\label{nuclei}
\subsection{The Milky Way Nuclear Star Cluster}

The supermassive black hole (SMBH) at the centre of our Galaxy (M$_{\rm{SMBH}}\sim 4 \cdot 10^{6}$ M$_{\odot}$) is surrounded by a massive, very compact star cluster which is commonly referred to as the Milky Way nuclear stellar cluster (MWNSC), with structural properties that are reminiscent of a massive globular cluster. 
Due to its proximity, such a star cluster provides us with a wealth of data which are not available for extragalactic nuclei. 

More generally, NSCs are very massive (M$_{\rm{NSC}}\sim 10^{6-7}$ M$_{\odot}$) and compact objects (radius $R_{\rm{eff}}\sim 2-5$ pc), and the Milky Way NSC offers a representative example (M$_{\rm{MWNSC}}\sim 2.1 \cdot 10^{7}$ M$_{\odot}$ and $R_{\rm{MWNSC}}^{\rm{eff}}\sim 4.2$ pc, see \citealt{scho2015}). Interestingly, NSCs host both old and young stellar populations, and spectroscopy is needed in order to reliably differentiate them (Kacharov et al., in prep). Two main formation scenarios have been suggested: (1) gas infall and in-situ star formation \citep{bekki,milos2004}; (2) accretion of GCs drifted into the Galactic centre via dynamical friction \citep{tremaine76,rcd1993,antonini2012}. On one hand, the young stellar population of the NSC has a density profile which is inconsistent with infalling cluster hypothesis \citep{feldmeier15}. 

On the other hand, $N$-body simulations of the NSC formation from cluster infall give as result a system which is very similar to the MWNSC (similar mass, central BH mass, flattening $\sim$0.7, density profile and kinematic properties, see \citealt{Tsatsi2017}). The observational characterization and dynamical interpretation of kinematical substructures can help shedding light on the discrepancy between these two models  \citep{feldmeier15,Cole2017}. 

Furthermore, the MWNSC shows a spread in metallicity distribution \citep[e.g., see][]{Feldmeier-Krause2017}. The radial distribution of low metallicity stars is consistent with the one of higher metallicity \citep{Do17}. The observed stellar metallicity in the MWNSC indicates that only a small fraction of stars have metallicity close to the one typical of GCs \citep{Do2015}, which suggests that in-situ star formation must have occurred. On the basis of these chemical and kinematic constraints, it appears that, at least for the MWNSC, the two formation channels are not mutually exclusive.

\subsection{Galactic centre objects: G1 \& G2} 
Since its discovery \citep{Gillessen12}, the gas and dust cloud G2 orbiting Sgr A* has caught the attention of the astronomical community. The recently recognized objects G1 and G2 are self-luminous objects with cold dust photospheres. They follow eccentric orbits around the SMBH at the Galactic Centre, and come close enough at periapse to suffer tidal removal of some of their extended material. Given that both G1 and G2 have survived their periapse passages, it has been strongly suggested that they are stellar objects \cite[e.g., see][]{Witzel2014} rather than compact dust clouds \cite[e.g., see][]{Pfuhl2015}, but their nature is still heavily debated.

On one hand these objects could be the result of relatively recent stellar mergers of binary pairs \citep{stephan16} whose internal orbits underwent rapid evolution toward extremely high eccentricities in response to the gravitational influence of a third body, the Galactic SMBH. The mergers should equilibrate to normal single stars on a Kelvin-Helmholtz time scale, but such induced mergers should be sufficiently common that other G-type sources are likely present.
On the other hand a compact source model, where G2 is the outflow from a mass-losing star, has been presented \citep{ballone17}. By means of 3D adaptive mesh refinement hydrodynamical simulations with the grid code PLUTO\footnote{\url{http://plutocode.ph.unito.it/}}, this work provided a more detailed and realistic comparison to the observed position-velocity diagrams, showing that a slow (i.e., 50 km/s) outflow, with parameters roughly similar to those of a T Tauri star,
can reproduce G2, and that in few years the central source should decouple from the previously ejected material.

\subsection{Nuclear star clusters in external galaxies}

Nuclear star clusters are a common feature in the majority of early- and late-type galaxies, with occupation fractions of $\, \gtrsim 60 - 70$\% \citep[e.g.][]{durell_1997, carollo_1998,boeker_2002,boeker_2004}. As for SMBHs, their masses correlate well with other properties of their host galaxies (such as velocity dispersion and total mass) possibly indicating a common evolutionary process for the host galaxy, the NSC, and the central BH; for this reason, often they are indistinctively referred to as Central Massive Objects \citep[see e.g.][]{ferrarese_2006}. 

\citet{georgiev_2016} explored scaling relations between NSC mass and host galaxy total stellar mass using a large sample of NSCs in late- and early type galaxies, including those hosting a central BH. They found differences in the slope of the correlation between NSCs in late-type and early-type galaxies among other trends depending on the galaxy type. This indicates possible different (ongoing) evolutionary processes in NSCs, depending on the host galaxy type. 

\section{Blue stragglers / stellar collisions and their products}
% Maxwell Cai
% Antonio Sollima
% Nora Luetzgendorf
%heavily edited by ALV
\label{collisions}
\subsection{Detection of blue stragglers}
Collisions involving unevolved main sequence (MS) stars or mass transfer in binaries with MS secondary components could lead to a rejuvenation of the resulting object \citep{McCrea1964}, which will appear brighter and warmer than other MS stars (blue straggler stars; BSS).

From the observational point of view, the need of conducting a systematic analysis of the BSS population of GCs in ultraviolet bands, where they emit the largest fraction of their light, is emerging. This is particularly true in dense environments, where crowding effects reduce  the photometric completeness of faint sources. For this reason, the contribution of the HST UV Legacy Survey \citep{Piotto2004}, thanks to the combined effect of high-resolution and UV sensitivity, appears as the most promising dataset to explore this population in the centre of Galactic GCs. However, future facilities providing similar performances will be needed in the near future when HST will end its mission. 

While the main formation channels of BSS have been clearly identified \citep{McCrea1964} and their efficiency as a function of the main cluster parameters has been both explored observationally \citep{Piotto2004} and interpreted in theoretical studies \citep{Davies2004}, still other observational evidence requires a proper interpretation. In particular, the evidence of parallel BSS sequences in the colour-magnitude diagram of M30 and a few other clusters \citep{Ferraro2009, Dalessandro2013} still awaits for a comprehensive explanation (i.e., can BSS formed through different formation channels contribute to produce the claimed bimodality? see e.g. \citealt{Xin2015, Jiang2017}). 

\subsection{Blue stragglers as dynamical clocks}
BSS are more than just exotic objects in a star cluster: their unique mass range and time scale of formation make them an ideal tool to study the dynamical state of their host systems. As a result of mass segregation process, BSS are often more centrally concentrated than other GC stars. Computing the radial distribution of BSSs in clusters with different dynamical ages, \citet{ferraro_2012} found  that clusters can be categorised in three families based on their radial distribution profiles of BSSs. For the intermediate-age GCs (family II) the BSS distribution exhibits a minimum at position $r_{\rm{min}}$ which is directly correlated with the clusters relaxation time scales, making BSS distributions ideal observables for the cluster's dynamical age. However, a tension between observation and theory has been put forward regarding this bimodal radial distribution which is not expected according to extensive surveys of Monte Carlo simulations \citep{Hypki2017}.

Recently, \citet{alessandrini_2016} and \citet{lanzoni_2016} suggested $A+$ (i.e., the area between the cumulative radial distribution of BSS and that of a reference population) as a dynamical observable. This high-signal feature is easy to measure and does not require binning, making it a more stable observable than the radial distribution profile . $A+$ also shows the same strong correlation with the cluster's dynamical time scales \citep{Ferraro2018bs}. 

\subsection{Simulations of stellar collisions}
From the computational perspective, enormous progresses have been made in the last years in the simulation of close interactions in star clusters. Still, direct integration of three-four body interactions, which are crucial to study the formation of such exotic species, severely slow down performances of all the $N$-body and Monte Carlo codes. This can constitute a bottleneck in particular when rare events (like BBH mergers and interactions between neutron stars) require simulations of real-size ($N>10^{6}$) clusters. 

Furthermore, while simple recipes are already included in all simulation codes \citep{Hurley2001, 2015MNRAS.451.4086S}, the complex interplay between stellar and dynamical evolution is still far from being properly accounted with a fully consistent SPH-stellar evolution approach, in particular when binary stars are
considered. Nevertheless, the development of computational environments where different codes simultaneously interact, allows today to move the first steps in facing complex collisional processes where $N$-body, gas accretion and UV radiation must be taken into account, e.g. to model the formation of very massive stars in the first stellar complexes \citep{Hosokawa2012} or the evolution of triples in a common Roche lobe overflow phase \citep{deVries2014}. 

\section{Compact objects in star clusters}
% Sara Rastello
% Alice Zocchi
% Nora Luetzgendorf
%heavily edited by ALV
\label{comp_objects}

\subsection{Binaries in star clusters}

X-ray sources are good tracers of compact binaries, especially low mass X-ray binaries (LMXBs), that are systems in which a compact object, i.e., a white dwarf (WD), a neutron star (NS) or a BH, accretes matter from a low-mass companion star. Accretion occurs through Roche-lobe overflow and disk formation around a compact object, or, in the case of red giants, wind-fed accretion partially captured by the compact object. NS binaries are mainly millisecond pulsars (MSPs), which form in dense clusters and are subsequently ejected as a consequence of tidal disruption and evaporation of the cluster \citep[see e.g.][]{frag_pav_ban}. WD binaries, (WD-WD or WD-MS stars) can be found in nearby GCs. 

In particular, WD-MS binaries are known as cataclysmic variables (CVs). The optimal way to identify possible CVs in GCs is by combining different techniques, to measure their optical variability, blue colour, H$\alpha$ excess, and X-ray emission \citep{knigge11}. \cite{Belloni2016} analysed the population of CVs in a sample of 12 GCs evolved with MOCCA, by considering two initial binary populations. They found that a population of CVs is mainly found in later stages of the evolution of GCs, and that selection effects can drastically limit the number of observable CVs. They also found that the probability of observing CVs during the outburst is extremely small, and they concluded that the best way of detecting such objects is by searching for variabilities during the quiescent phase. In addition, magnetic fields might be needed to explain the rare frequency of outbursts amongst bright CVs ($\sim 10 \%$).

\subsection{Stellar-mass black holes}
A relevant aspect of the dynamics of stellar-mass BHs in dense stellar environments is related to the retention of massive objects in star clusters. Indeed, the retention of BHs (and NSs) can have a strong influence on the global evolution of globular clusters \citep[e.g., see][]{breenheggie13,2015MNRAS.449L.100C,pav18}. Initial mass function and BH formation mechanisms (especially kicks) play a major role in determining the subsequent evolution of the BH population in a GC \citep{chat2016,mandel16}, and a clear distinction is seen between the mass loss due to stellar evolution \citep[connected with metallicity,][]{bhspera} and relaxation. BH subsystems can be formed and preserved in dense environments, provided the cluster has a sufficiently long relaxation time \citep{breenheggie13,breenheggie13b}. 

Massive binary black holes (BBHs) are preferentially formed in low-metallicity and dynamically active stellar environment \cite[e.g., see][]{2016MNRAS.459.3432M}. Recent works carried out with Monte Carlo simulations pointed out that more massive clusters are more likely to trigger BBH mergers \citep{rod2016,rod2016b} although these events may also take place in stellar systems with a lower density (such as open clusters, see Section~\ref{grav_waves}). The recent detections of gravitational waves from merging BBH have the potential to revolutionize our understanding of compact object astrophysics, but to fully utilize this new window into the universe, these observations must be compared to detailed theoretical models of BBH formation throughout cosmic time. {\cite{Tanikawa2013} and, more recently, \cite{Fujii2017} calculated the detection rates of gravitational waves emitted from merging BBHs in star clusters modelled as direct N-body systems. \citet{Hurley2016} have also reported quantitative confirmation of the merging of two stellar-mass BHs in a binary system which was dynamically formed in a moderately-sized direct N-body model. Merger rates determined on the basis of Monte Carlo modelling approaches have also been intensively explored \citep[e.g., see][]{rod2016,Askar2017,Hong2018}. More details are discussed in Section~\ref{grav_waves}.}

\subsection{Intermediate-mass black holes}

Intermediate-mass black holes (IMBHs) are defined as covering a mass range of $10^2 - 10^5$ M$_{\odot}$ and have become a promising field of research. With their existence it could be possible to explain the rapid growth of SMBHs, which are observed at high redshifts \citep{2006NewAR..50..665F}, by assuming that IMBHs act as SMBH seeds \citep[e.g.,][]{2001ApJ...562L..19E,2009ApJ...696.1798T}. Recent discoveries of black holes in the centres of dwarf galaxies \citep{2015ApJ...813...82R} have shown that the mass range between supermassive and stellar-mass black holes is by far not empty. However, whether or not IMBHs exist in ordinary GCs is still under debate.

Several formation scenarios have been proposed for these objects, but conclusive evidence on which one is preferred is still missing. \citet{2001ApJ...551L..27M} proposed that IMBHs could be the remnants of Population III stars, obtained after an evolution of a few Myr. However, if this is indeed the route to build up IMBHs, we should not expect them to be found in clusters of Population I stars. \citet{2004Natur.428..724P} suggested that an IMBH could be the end product of a runaway collision in the centre of star clusters. This process needs specific initial conditions, and requires the time scale of the mass segregation of the most massive stars to be shorter than the evolution time-scale for those stars, to avoid them to evolve before they start to collide.
Recently, another scenario has been proposed, also  indicating that these objects form in star clusters. \citet{2015MNRAS.454.3150G} proposed that an IMBH is formed as a consequence of the build-up of BH mass due to mergers in dynamical interactions and mass transfers in binaries; in spite of the ones described before, this scenario does not require the onset of particular initial conditions, but the process of IMBH formation is highly stochastic. A larger formation probability is obtained for clusters with larger concentration, and for earlier and faster BH mass build-up. A great effort has been devoted to numerical simulations, not only to test these formation scenarios, but also to understand which properties of the host system are mainly determining the presence and the mass of an IMBH in their centre. 

Direct detection of an IMBH is extremely challenging because GCs are almost gas free. Radio observations \citep{2012ApJ...750L..27S} of the cores of three Galactic GCs (M15, M19 and M22) do not unveil any compact source: this non-detection sets the upper limit of the masses of IMBHs in these systems to $\sim 3-9 \times 10^2$ M$_{\odot}$. Such a result has been recently confirmed by a more extensive radio continuum survey conducted on Galactic GCs \citep{Tremou2018}. These limits suggest that either IMBHs more massive than $10^3$ M$_{\odot}$ are rare in GCs, or that if they are present, they accrete in a very inefficient manner. On the other hand, an X-ray outburst has been detected in a star cluster located off-center of a large lenticular galaxy; such an event has been interpreted as providing strong evidence that the source contains an IMBH of $10^4$ M$_{\odot}$. 
\citet{2016MNRAS.460.2025K} proposed a different method to look for IMBHs in GCs, by means of gravitational microlensing. From a suite of simulations, they estimate the probabilities of detecting such an event for Galactic GCs: as an example, they consider M22, and conclude that if it hosts an IMBH with mass $10^5$ M$_{\odot}$, there is a probability of 86\% of detecting an astrometric microlensing event over a baseline of 20 yr.

Nevertheless, even though direct detections of IMBHs are challenging, signatures of the presence of an IMBH are imprinted in the phase-space distribution of stars in its immediate surrounding \citep{1976ApJ...209..214B}. {The effects of the presence of an IMBH on the structural and kinematic properties of the host star clusters have subsequently been explored in detail by means of direct N-body simulations \citep{Baumgardt2004a,Baumgardt2004b,Baumgardt2005}.} In particular, two signatures are at the basis of the observational claims for the detection of IMBHs in Galactic GCs: the detection of a shallow cusp in the surface brightness profile \citep[e.g., see][]{2008ApJ...676.1008N} and the presence of a rise in the projected velocity dispersion profile towards the centre \citep[e.g., see][]{2010ApJ...710.1032A,2013A&A...552A..49L}. However, these signatures can also be due to other processes: core collapse, mass segregation, or a population of binaries in the centre also produce a shallow cusp in the brightness profile, as shown with dedicated $N$-body simulations \citep{2010ApJ...720L.179V}, and the central rise in the velocity dispersion profile is also not unique \citep{2017MNRAS.468.4429Z, Zocchi2018}. The controversial case of $\omega$ Cen is an example of this degeneracy: isotropic spherical models only reproduce the observed rise in the projected velocity dispersion when a central IMBH of mass $\sim 4\times 10^4$ M$_{\odot}$ is included \citep{2008ApJ...676.1008N}, while by comparing anisotropic models (with radial anisotropy in the core and tangential anisotropy in the outer parts) to proper motion measurements an upper limit to the IMBH mass of only $\sim 7 \times 10^3$ M$_{\odot}$ is obtained \citep{2010ApJ...710.1063V}.

This degeneracy in the signatures of the presence of an IMBH has also been explored by means of numerical simulations. \citet{2013A&A...558A.117L} presented a set of direct N-body simulations of GCs in an external tidal field, considering several values for IMBH masses, BHs retention fractions, and binary fractions. Their results show that the presence of an IMBH, or of a central population of binaries or stellar-mass BHs increases the escape rate of high-mass stars; these simulations show a good agreement with observational mass functions and structural parameters of GCs. A similar result has been found by \citet{2016MNRAS.455...35A}, who proposed an analysis of numerical simulations showing that the excess of mass in the centre of a cluster could be due to the presence of a subsystem of heavy remnants orbitally segregated, instead of being due to an IMBH (see also  \citealt{Zocchi2018}). {Finally, the coexistence of an IMBH and of a population of stellar-mass BHs has been explored by \cite{Leigh2014}, by means of direct N-body simulations.}

In addition, some controversy has recently arisen when comparing data obtained by means of integrated light spectroscopy to measurements of line-of-sight velocities of single stars \citep[see for example the emblematic case of NGC 6388,][]{2011A&A...533A..36L,2013ApJ...769..107L,2015A&A...581A...1L}. Studies of mock observations of numerical simulations of star clusters with and without a central IMBH have been carried out to determine the magnitude of this effect. \citet{2015MNRAS.453..365B} showed that luminosity-weighted IFU observations can be strongly biased by a few bright stars introducing a scatter in the measurement of the velocity dispersion up to $\sim 40\%$ around the expected value: this prevents any sound assessment of the central kinematics, and does not allow for an interpretation of the data in terms of the presence of a central IMBH. \citet{2017MNRAS.467.4057D} estimated that in $20 \%$ of the cases the analysis of data did not allow for a statistically significant detection of IMBHs having mass equal to $3\%$ of the mass of their host, because of shot noise due to bright stars close to the IMBH, and that when considering a smaller fractional mass for the IMBH ($\sim 0.1$) the rate of non-detections corresponds to $75 \%$ of the cases. The combination of data from new-generation facilities (such as Gaia proper motions and MUSE spectroscopic data of the central regions of clusters) will provide further constraints to the GC dynamics, allowing us to determine if IMBHs are hiding in the cores of GCs, and how massive they are.

Leaving the uncertainties and difficulties of detecting IMBHs in GCs aside, other systems have proven to host IMBHs at their centre. In the last decade, detections of AGN in nearby dwarf galaxies have provided great evidence of the existence of SMBHs in the lower mass regime \citep[e.g.][]{2015ApJ...809L..14B, 2011Natur.470...66R,2003ApJ...588L..13F, 2004ApJ...607...90B, 2015ApJ...809..101D}. Inspired by those recent discoveries, researchers have branched out to search for BHs in smaller stellar systems such as ultra compact dwarf galaxies \citep{2014Natur.513..398S} and even our closest neighbour, the Large Magellanic Cloud \citep{2017ApJ...846...14B}. 

{In addition, there have been recent suggestions of the presence of candidate IMBHs embedded in gas clouds in the region of our Galactic Centre, as based on emissions in the millimetre \citep{Oka2016, Oka2017} and infra-red bands \citep{Tsuboi2017}. More generally in this context, a powerful diagnostic tool is offered by the phenomenology of tidal disruption events. Indeed, there have been recent reports of events that may involve IMBHs \citep{Lin2018, Kuin2018}, with fresh developments also on the theoretical side \citep[e.g., see][]{Rosswog2009,Tanikawa2018,Anninos2018}.} 

Over the next years, {this spectrum of studies will greatly} contribute in filling the lower mass range of the BH mass/host galaxy property relations and enhance our understanding of BH evolution and occupation fractions.

\section{Astrophysical sources of gravitational waves}
% Václav Pavlík
% Sara Rastello
% Maxwell Cai
% heavily edited and completed by ALV
\label{grav_waves}

Currently, the detection by LIGO of GWs emitted by systems identified as coalescing binary black holes (BBH) suggests that these systems are primary sources of GW emission \citep{2016PhRvL.116f1102A, 2017PhRvL.118v1101A, GW170608, GW170814}. This long-awaited empirical evidence has immediately sparked a debate of great theoretical importance to understand the origins of these BBH systems and to predict their frequencies over cosmic time. So far, two main formation channels have been proposed: dynamical interactions in dense stellar environment \citep{PortegiesZwartMcMillan2000} and evolution of isolated stellar binaries \citep{Belczynski2002}.

Stellar environments may also play an important role in shaping the properties of the merger. Using three-body scattering experiments to model the scattering process between stars and BBHs, \citet{2017PhRvD..95h4032R} suggest that galactic nuclei may harbour supermassive BBHs. Systems in that mass class are expected to be primary sources for GW emission, especially in the detection range relevant for the upcoming LISA observatory. Similarly, \citet{Antonini2016} and \citet{2017ApJ...835..165B} suggest that the densest population of stellar-mass BHs is expected to be found in galactic nuclei. A fraction ($\sim 30\%$) of these BHs can reside in binaries.

An extensive survey of simulations, carried out with the MOCCA code (see Section \ref{numerical_methods}), followed the long-term evolution of GCs, assessed the retention of their stellar-mass BHs, and allowed to estimate the local merger rate densities for the BBHs originating from GCs \citep{Askar2017, Hong2018}. {Detection rates of BBH mergers have been determined also on the basis of direct N-body models \citep{Tanikawa2013,Fujii2017}.}
In other models with an IMBH, two-body collisions between an IMBH and a stellar mass BH are very frequent \citep{Leigh2014, Morawski2018}. 

Although open clusters and young massive clusters, in general, produce less binary BH mergers per cluster than the GCs, they are more numerous, which makes them a competitive formation environment \citep{sambaran1}. By means of $N$-body simulations carried out with the code NBODY7, \cite{sambaran2} studied the evolution of BBH populations in young massive and starburst clusters estimating the merger rate for each model. Their overall contribution to the observable BBH inspiral rate in the universe could be at least comparable to that from classical GCs, thereby potentially favouring the BBH detection rate from the dynamical channel. 
In addition, this study shows a tendency of the BBHs to coalesce within the clusters. In particular, the general-relativistic BBH mergers continue to be mostly mediated by triples that are bound to the clusters, rather than happen among the ejected BBHs.
In fact, the number of such in-situ BBH mergers, per cluster, tends to increase with the introduction of a small population of primordial stellar binaries.

These BBH mergers are well consistent with the LIGO detection window and suggest a full-sensitivity LIGO detection rate of up to hundreds of BBH mergers from stellar clusters, per year.
Rastello et al. (in prep.) also studied, by means of $N$-body simulations carried out with NBODY7, the evolution of massive BBHs (of the class expected to be detectable by LIGO) in open clusters ($10^{2} - 10^{3}$ M$_{\odot}$), deriving a merger rate of the order of 2.1 yr$^{-1}$ Gpc$^{-3}$. { Most importantly, it has been recently shown that general relativistic corrections are crucial to appropriately assess the rates of mergers of BBH on eccentric orbits \citep{Samsing2017,Samsing2018}.}

Concerning the origin of other specific LIGO events, \cite{2017arXiv170607053B} argues that GW170104 \citep{2017PhRvL.118v1101A} may be formed via classical isolated binary evolution. The DRAGON simulations \citep{Wang2016}, which follow the dynamical evolution of a globular cluster ($N=10^6$) using state-of-the-art high-performance GPUs and post-Newtonian treatment for BH binaries, found black hole mergers producing waveforms similar to GW150914 \citep{2016PhRvL.116f1102A}. The results of \cite{2017MNRAS.469.4665P}, who also performed direct $N$-body simulations, also suggest that GW150914 is the consequence of a dynamically formed BBH. We are, therefore, just scratching the surface of the theoretical implications of the first LIGO detections.

\section{Conclusions}
% Anna Lisa Varri
\label{concl}

With their elegant analogy to the classical gravitational $N$-body problem and tantalizing access to their resolved stellar populations, star clusters have been a prime target for both theoretical and observational astronomers. Resting on such a distinguished legacy, they have often been considered as a ``solved problem'' in astronomical research, but, in fact, a series of recent discoveries about their chemical, kinematic and structural properties have revolutionised our traditional picture of star clusters, making them baffling chemodynamical puzzles, prolific black hole cradles, precious galactic beacons, and even unexpected tools for galaxy evolution and near-field cosmology. 

Now, standing at the crossroads of Gaia’s blooming era of ``precision astrometry'' and LIGO’s revolutionary beginnings of ``gravitational wave astronomy'', the rich internal dynamics of star clusters truly brings them back to the centre stage, and the surprises are far from over.     

We leave to the interested reader to compare our current
understanding of { these topics} with the open questions (and their envisaged timeline) identified more than a decade ago by \citet{Sum_M6} and to assess the progress made in each { frontier} area. Many milestones have been achieved, especially on the computational side, thanks to the development of numerical algorithms which enabled us to finally ``solve'' the gravitational $N$-body problem \citep{Aarseth2003,Nitadori2012,Wang2016,Heggie2014}, at least in the sense of following the entire evolution of models of individual clusters, with a realistic number of particles and fundamental internal and external effects. One of the original goals of providing a numerical framework for realistic simulations of the many aspects of the evolution of dense stellar systems
\footnote{\url{http://www.manybody.org/modest/}} is about to be fulfilled \citep{AMUSE}. But many new challenges have also emerged, mostly driven by the outstanding progress on the observational side. We do not dare providing a comprehensive list, but we nonetheless  wish to conclude by highlighting some pressing questions which have emerged during the 17th annual meeting of the MODEST community: 
\vspace{0.1cm}\\
\begin{itemize}
\item How can we leverage the emerging phase space richness of old globular clusters to decipher the dynamical signatures of their formation and evolutionary processes?   
\item Which is the best numerical approach to efficiently, yet accurately, attack the ``post-million body problem''?
\item How can we explore, from first principles, the multi-scale, multi-physics problem of the formation of globular clusters in a cosmological context?
\item What is the role of star cluster environments in shaping the dynamical evolution of planetary systems?
\item How can we bridge the gap between the current knowledge of star formation processes, the origin of the first stellar aggregates, and their subsequent long-term evolution?
\item How can we exploit at best the synergy between Gaia, LIGO, JWST and other upcoming facilities to finally unveil when, where, and how globular clusters formed? 
\item What is the physical origin of multiple stellar populations in globular clusters? 
\item What is the nature of the complex nexus between globular and nuclear star clusters and their formation channels?
\item How can we observationally characterise and dynamically interpret the role of dark remnants in the evolution of collisional systems? 
\item Do globular clusters really harbour intermediate-mass black holes, and, in such a case, how did they form?  
\item What is the origin of the stellar-mass black holes binary revealed by the first LIGO events?
\end{itemize}
\vspace{0.3cm}

The MODEST-17 conference has provided evidence that the community invested in ``Modelling and Observing DEnse STellar systems'' is creative, active, collaborative and diverse. Since its early years, such a group of researchers has much evolved, in size and scope, but the overarching goal of continuing to explore the richness of collisional stellar dynamics with a multi-disciplinary approach is more timely than ever. As the broader astrophysical community is progressively appreciating the many roles played by star clusters throughout cosmic time (e.g., as contributors to the reionisation epoch, tracers of galaxy formation, cradles of gravitational wave sources), an additional challenge for the MODEST collaboration will be to find {new} synergies with { appropriate} theoretical and observational research communities, in {order} to formulate a comprehensive, modern view of these fascinating stellar systems in the evolving landscape of contemporary astrophysics.

%%%%%%%%%%%%%%%%%%%%%%%%%%%%%%%%%%%%%%%%%%%%%%
%%                                          %%
%% Backmatter begins here                   %%
%%                                          %%
%%%%%%%%%%%%%%%%%%%%%%%%%%%%%%%%%%%%%%%%%%%%%%
\section*{List of abbreviations}

\begin{itemize}
\addtolength{\itemindent}{-0.9cm}
\item [] AMUSE - Astrophysical Multipurpose Software Environment
\item [] ATHENA - Advanced Telescope for High-ENergy Astrophysics
\item [] BBH - Binary Black Hole
\item [] BH - Black Hole
\item [] BSS - Blue Straggler Star
\item [] CGM - Circum-Galactic Medium
\item [] CMC - Cluster Monte Carlo 
\item [] CMD - Color Magnitude Diagram
\item [] CPU - Central Processing Unit
\item [] CV - Cataclismic Variable
\item [] DR - Data Release
\item [] EAGLE - Evolution and Assembly of GaLaxies and their Environments
\item [] E-MOSAIC - MOdelling Star cluster population Assembly In Cosmological Simulations within EAGLE
\item [] ESA - European Space Agency 
\item [] ESO - European Southern Observatory
\item [] FIRE - Feedback In Realistic Environments 
\item [] GC - Globular Cluster
\item [] GMT - Giant Magellan Telescope
\item [] GPU - Graphics Processing Unit
\item [] GW - Gravitational Waves
\item [] HST - Hubble Space Telescope
\item [] IFU - Integrated Field Unit
\item [] IMBH - Intermediate-Mass Black Hole
\item [] IMF -  Initial Mass Function
\item [] IGM - Inter-Galactic Medium
\item [] ISM - Inter-Stellar Medium
\item [] JWST - James Webb Space Telescope
\item [] KAGRA - KAmioka GRAvitational Wave Detector
\item [] LAMOST - Large Sky Area Multi-Object Fibre Spectroscopic Telescope
\item [] LIGO -  Laser Interferometer Gravitational-Wave Observatory 
\item [] LISA - Laser Interferometer Space Antenna
\item [] LMXB - Low-Mass X-ray Binaries
\item [] M/L - Mass-to-light ratio
\item [] MYStIX  -Massive Young Star-Forming Complex Study in Infrared and X-ray
\item [] MF -  Mass Function
\item [] MOCCA - MOnte Carlo Cluster simulAtor
\item[] MODEST - Modeling and Observing DEnse STellar systems
\item [] MPI - Message Passing Interface
\item [] MS - Main Sequence
\item [] MSP - Milli-Second Pulsar
\item [] MUSE - Multi-Unit Spectroscopic Explorer
\item [] MW - Milky Way
\item [] MWNSC - Milky Way Nuclear Star Cluster
\item [] NASA - National Aeronautics and Space Administration
\item [] NS - Neutron Star
\item [] NSC - Nuclear Star Clusters 
\item [] ONC - Orion Nebula Cloud 
\item [] P$^3$T - Particle-Particle Particle-Tree
\item [] RAVE - RAdial Velocity Experiment
\item [] SPARC-SC - Symplectic Particle tree and Algorithmic Regularization Code for Star Clusters
\item [] SDSS - Sloan Digital Sky Survey
\item [] SIMD - Single Instruction Multiple Data Extensions
\item [] SMBH - Super-Massive Black Hole
\item [] SSE - Streaming SIMD Extensions
\item [] TGAS - Tycho-Gaia Astrometric Solution
\item [] UCD -  Ultra-Compact Dwarf Galaxy
\item [] UV - Ultra Violet
\item [] WD - White Dwarf
%\item [] 
\end{itemize}
\newpage

\section*{Declarations}
\begin{backmatter}

% \section*{Competing interests}
%   The authors declare that they have no competing interests.

\section*{Availability of data and material}
Data sharing not applicable to this article as no datasets were generated or analysed during the current study.

\section*{Competing interests}
The authors declare that they have no competing interests.

\section*{Funding}
ALV acknowledges support from a Marie Sklodowska-Curie Fellowship (MSCA-IF-EFRI NESSY 658088) {and the Institute for Astronomy at the University of Edinburgh}; MXC from the Poles Programme (initiated by the Belgian Science Policy Office, IAP P7/08 CHARM) and the European Union’s Horizon 2020 research and innovation programme under grant agreement No. 671564 (COMPAT project); FD from the DFG Priority Programme 1573 ``The physics of
the interstellar medium''; VP from Charles University, grants SVV-260441 and GAUK-186216; SR from Sapienza, University of Rome, grant AR11715C7F89F177; AZ from Royal Society (Newton International Fellowship Follow-up Funding). These funding bodies had no direct role in the design of the study and collection, analysis, and interpretation of data and in writing the manuscript.

\section*{Author's contributions}
%     Text for this section \ldots
All authors are equal contributors, hence are listed in alphabetical order; ALV coordinated the manuscript writing.

\section*{Acknowledgements}
We are most grateful to Ladislav \v{S}ubr, Jaroslav Haas and all SOC and LOC members for an excellent conference programme, engaging scientific environment, and their warm welcome to Prague. Support from Charles University, Prague, where MODEST-17 was held, is also gratefully acknowledged. We wish to thank especially Simon Portegies Zwart and Ladislav \v{S}ubr for proposing and providing stimulus and { support} to the { publication} this review, and Melvyn Davies, Francesco Ferraro, Rainer Spurzem for their early encouragement. The title of the manuscript has been suggested by Simon Portegies Zwart. {We are very grateful to the Referees, Douglas Heggie and two anonymous researchers, for taking the time to read the manuscript, offer accurate comments and suggest pertinent references, as well as the Editor, Eiichiro Kokubo, for overseeing the review process.} We acknowledge the input provided by all MODEST-17 participants, and, more generally, by the members of the MODEST community. While preparing this manuscript, we made intensive use of NASA’s Astrophysics Data System and the arXiv e-print repository.

%%  After submission of the .TEX file,                     %%
%%  you will be prompted to submit your .BBL file.         %%

% if your bibliography is in bibtex format, use those commands:
\bibliographystyle{spbasic} % Style BST file (bmc-mathphys, vancouver, spbasic).
\bibliography{bmc_article}      % Bibliography file (usually '*.bib' )

\end{backmatter}
\end{document}